\documentclass[twocolumn,english,aps,prb,showpacs]{revtex4-1}
\usepackage[T1]{fontenc}
\usepackage[latin9]{inputenc}
\usepackage{mathrsfs}
\usepackage{amssymb}
\usepackage{graphicx}
\usepackage{wasysym}
\usepackage{esint}

\makeatletter
\usepackage{babel}

\usepackage{babel}

\makeatother

\usepackage{babel}
\begin{document}

\title{Nonequilibrium charge transport through Falicov-Kimball structures
connected to metallic leads}

\author{Martin \v{Z}onda}

\address{Institute of Physics, Albert-Ludwig University of Freiburg, Hermann-Herder-Strasse
3, 791 04 Freiburg, Germany}

\author{Michael Thoss}

\address{Institute of Physics, Albert-Ludwig University of Freiburg, Hermann-Herder-Strasse
3, 791 04 Freiburg, Germany}
\begin{abstract}
Employing a combination of a sign-problem-free Monte Carlo approach
and nonequilibrium Green's-function techniques, we study nonequilibrium
charge transport in a model heterostructure, where a two-dimensional
spinless Falicov-Kimball system is coupled to two noninteracting leads.
We show that the transport characteristic depends sensitively on the
electrostatic potential in the system and exhibits different properties
for different phases of the Falicov-Kimball model. In particular,
pronounced step-like changes of the current and transmission are observed
at the phase boundaries, evident even on a logarithmic scale. Analyzing
finite size effects, we find that with the method used a relatively
small system can be utilized to address specific thermodynamic limits.
Phenomenon of the localization is discussed as well.
\end{abstract}
\maketitle

\section{Introduction}

The study of strongly correlated electron systems (SCESs) has been
a field of intensive research for several decades. In recent years,
the investigation of out-of-equilibrium phenomena has received particular
attention. Experimental studies in this context comprise a variety
of different techniques and architectures including time-dependent
problems in quantum dots \cite{Wingreen1993,Weil2002,DiCarlo2003,Watson2003,Kohler2005,Leek2005,Hohls2012,Harabula2017},
molecular bridges and nano-wires \cite{Galperin2008,Zimbovskaya2011,Ballmann2012,Ferdinand2017,Bouvron2018,Burzuri2014},
layered systems, junctions, and hetero-structures \cite{Gariglio2002,Ohtomo2002,Bozovic2004,Baiutti2018},
as well as dynamics of quenched cold atoms in optical latices \cite{Greiner2002,Greiner2002b,Sadler2006,Clos2016}
or ultrafast electronics \cite{Cavalleri2001,Kubler2007,Liu2012}
and pump-probe experiments on the charge density wave materials (CDW) \cite{Clerc2007,Perfetti2008,Schmitt2011}.
Theoretical investigations have focused on open questions related,
e.g., to the formation of nonthermal steady state \cite{Rigol2007,EcksteinPRL2008},
dynamical phase transitions, and hidden metastable phases revealed
by driving \cite{Eckstein2009,Stojchevska2014,Vaskivskyi2015,Vaskivskyi2016,Reichhardt2018}
as well as to the evolution of open quantum systems in general \cite{Breuer-Petruccione,Breuer_RevModPhys2016}.

To study these open questions, a variety of theoretical methods have
been employed. Some of them are rather versatile, e.g., methods based
on nonequilibrium Green's functions \cite{Haug-Jauho,RammerRMP1986,Kamenev2009,Spicka2014};
however, they often rely on significant approximations, which limit
their validity to certain parameter regimes, such as, e.g., weak coupling
\cite{KadanoffBook,Lipavsky1986,Hermanns2012}. Other methods, e.g.,
nonequilibrium dynamical mean-field theory (DMFT) \cite{FreericksPRL2006,Aoki2014},
require so-called nonequilibrium impurity solvers \cite{Tanimura1989,MLMCTDH2003,Anders2005,Schollwoeck2006,Muhlbacher2008,GullRevModPhys2011,Hartle2013,Costi2014}.
Although there is rapid progress in the development of these solvers,
their application is still numerically very demanding. Especially
challenging is the study of long-time evolution including nonequilibrium
steady-state properties. It is, therefore, important to study special
cases of SCES models, which can be addressed by less demanding, but
still exact methods. The results of such studies not only help us
to gain a deeper understanding of nonequilibrium phenomena, but can
also serve as testing tools for addressing more challenging systems.
An example of such a model, which in recent years played an important
role in the studies of nonequilibrium SCES, is the Falicov-Kimball
model (FKM).

The spinless version of the FKM was initially introduced as a limiting
case of the Hubbard model, where one electron species is localized\cite{Hubbard63}.
Later, but independently and with spin degrees of freedom, the model
was used for the metal-insulator transitions in rare-earth and transition-metal
compounds\cite{FKM1969}. The FKM has been used successfully for the
description of numerous equilibrium phenomena (for overviews see Refs.~\onlinecite{GruberHPA1996,Freericks2004,FarkasovskyAPS2010,Cencarikova2011}),
for different properties of layered systems \cite{FreericksBook2006,Freericks2001,Hale2012,Kaneko2013}
and for problems related to transport, time evolution and nonequilibrium
steady states \cite{Turkowski2007,EcksteinPRL2008,Eckstein2009,Herrmann2016,Haldar2017,Qin2018,Haldar2016,Smith2017,Herrmann2018}.
From the conceptual point of view, the most profound advantage of
the FKM is the fact that it is exactly solvable in the limit of infinite
dimension (infinite coordination number) by means of DMFT \cite{FreericksRMP2003}
and that for finite lattices it can be addressed by an exact, sign-problem-free
Monte Carlo (MC) method \cite{MaskaPRB2006,ZondaSSC2009,Zonda2012,Huang2017}.
In addition, both these methods can be extended to nonequilibrium
without the necessity to introduce any approximation \cite{Turkowski2005,FreericksPRL2006,Aoki2014,Herrmann2018}.

In the present paper, we introduce a combination of the sign-problem-free
MC method with a nonequilibrium Green's-function technique, which
allows us to address the nonequilibrium steady state of a composite structure
consisting of a finite FKM system (or its various generalizations)
and infinite leads. An important advantage of this method is that
it can be applied to any lattice geometry and particle filling in
finite dimension. The only limiting factor is the total size of the
system. The method is therefore especially useful for lattices with
only few atomic layers in one or two directions (layered systems)
connected to semi-infinite leads \cite{ArrigoniPRL2013,FreericksBook2006}.
For such geometries, the method can describe large enough systems
to approach the thermodynamic limit.

Inspired by the recent rapid progress in understanding the equilibrium
and nonequilibrium properties of the two-dimensional (2D) FKM \cite{Antipov2016,Herrmann2016,Herrmann2018,Hohenadler2018}
we primarily concentrate on charge transport through finite square
FKM lattices driven by voltage differences of the lead potentials
(for illustration see Fig.~\ref{fig:Schema}). Our focus is on how
different phases of the FKM influence the transport characteristics and
how the nonequilibrium steady state influences the typical
phases of the FKM. Although, in the present paper, we primarily concentrate
on square lattices, we show that the results are also valid for
more general layered systems.

The rest of the paper is organized as follows. In Sec.~\ref{sec:Model-and-method},
we introduce the model and outline the methodology. In Sec.~\ref{sec:Results}
we discuss two different choices of the electrostatic potential in
the system. We analyze the equilibrium properties of the FKM necessary
for understanding the nonequilibrium steady state for the two choices
in Sec.~\ref{subsec:PhDF} and Sec.~\ref{subsec:PhDT}. In Secs.~\ref{subsec:TransportA},
\ref{subsec:TransportT} the respective transport properties of the
FKM are analyzed. We show that the potential has a crucial effect
on the transport properties and that the current is very different
in the ordered and disordered phase. The effects of the system size
are discussed in Secs.~\ref{subsec:SSEF}, \ref{subsec:SSET}. Sec.~\ref{sec:Conclusions}
concludes with a summary.

\section{Model and method\label{sec:Model-and-method}}

\begin{figure}
\includegraphics[width=1\columnwidth]{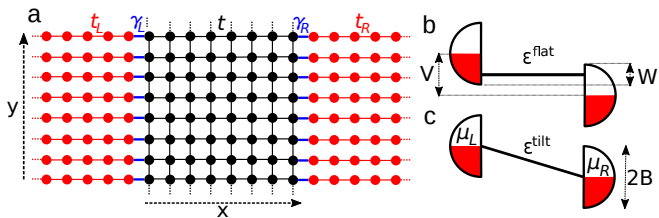}\caption{(a) Schematic picture of the heterostructure. The black part represents
the FKM system with nearest neighbor hopping $t$. The red parts are
noninteracting leads with hopping $t_{L,R}$ and the hybridization
interaction with hopping $\gamma_{L,R}$ is indicated in blue. (b-c)
Illustration of the introduced voltage bias with two limiting cases
of the potentials taken into account in the present paper. The elliptic
DOSs of the leads are shifted by $\epsilon_{L,R}$ and the condition
$\mu_{L,R}=\epsilon_{L,R}$ is used to keep the bands half filled
at any applied voltage. The system potential used is either (b) flat
$\varepsilon^{\mathrm{flat}}=-U/2$ or (c) tilted $\varepsilon^{\mathrm{tilt}}$(see
the text). \label{fig:Schema}}
\end{figure}

We consider a composite structure consisting of a large but finite
central system described by the spinless FKM and two infinite leads.
The Hamiltonian of this heterostructure reads $H=H_{S}+\sum_{l=L,R}H_{lead}^{l}+H_{hyb}^{l}$
(for illustration see Fig.~\ref{fig:Schema}(a)), where 
\begin{eqnarray}
H_{S} & = & -\sum_{i,j}t_{ij}d_{i}^{\dagger}d_{j}^{\phantom{\dagger}}+U\sum_{i}f_{i}^{\dagger}f_{i}^{\phantom{\dagger}}d_{i}^{\dagger}d_{i}^{\phantom{\dagger}}\nonumber \\
 &  & +\sum_{i}\varepsilon_{i}^{d}d_{i}^{\dagger}d_{i}^{\phantom{\dagger}}+\sum_{i}\varepsilon_{i}^{f}f_{i}^{\dagger}f_{i}^{\phantom{\dagger}},
\end{eqnarray}
represents the central system. 

Here, the first term describes spinless electrons moving on a lattice.
The geometry of the lattice and the hopping amplitudes are defined
by the hopping matrix $\mathbf{t}$. In the present paper, we focus
on square lattices with a constant nearest-neighbor hopping amplitude
$t$. We assume periodic boundary conditions at the edges perpendicular
to the system-lead interface. This mitigates the finite-size effects
in the $y$ direction. Most of the results presented below have been
obtained for lattice size $L=L_{x}\times L_{y}=20\times20$. Nevertheless,
the method can be used for any Hermitian matrix $\mathbf{t}$. The
second term represents a Coulomb-like local interaction between the
localized $f$ particle and itinerant $d$ electron on the same lattice
site. The last two terms describe the position-dependent potentials
that act on both particle species. This potential can be, in principle,
influenced by the leads and, depending on the interpretation of the
FKM, can differ for $f$ and $d$ particles \cite{MaskaPRL2008}.
However, throughout this paper we will assume that the $f$ particles
are also spinless electrons and are, therefore, affected by the potential
equally as $d$ electrons. Correspondingly, we set $\varepsilon_{i}^{f}=\varepsilon_{i}^{d}$.
This choice is advantageous with respect to the interpretation of
the FKM as a limiting case of the Hubbard model \cite{Hubbard63}.
Moreover, we fix the number of $f$ particles to $N_{f}=L/2$. This
reflects the fact, that only $d$ electrons are directly coupled to
the leads, which act as infinite reservoirs, and that the model
does not contain a hybridization between the two particle species.
The specific profiles of the potential are discussed in Sec.~\ref{sec:Results}.

The leads are taken to be noninteracting 
\begin{equation}
H_{\mathrm{lead}}^{l}=-t_{l}\sum_{\left\langle m,n\right\rangle }\left(c_{l,m}^{\dagger}c_{l,n}^{\phantom{\dagger}}+c_{l,n}^{\phantom{\dagger}}c_{l,m}^{\dagger}\right)+\epsilon_{l}\sum_{n}c_{l,n}^{\dagger}c_{l,n}^{\phantom{\dagger}},
\end{equation}
where $\left\langle m,n\right\rangle $ is a sum over the nearest-neighbor
sites, $t_{l}$ is the hopping for lead $l=L,\,R$ and $\epsilon_{l}$
represents an energy shift. Finally, the hybridization between lead
$l$ and the system is described by the term 
\begin{equation}
H_{\mathrm{hyb}}^{l}=-\gamma_{l}\left(\sum_{\left\langle i,n\right\rangle }c_{l,n}^{\dagger}d_{i}^{\phantom{\dagger}}+d_{i}^{\dagger}c_{l,n}^{\phantom{\dagger}}\right),
\end{equation}
with $\gamma_{l}$ being the corresponding hopping parameter. We assume
that the hybridization is turned on adiabatically starting in the
distant past with a decoupled state, where both leads and system had
been in thermal equilibrium. Note, that the initial conditions play
an important role because the FKM is an integrable model with the
occupations of the $f$ particles being integrals of motion \cite{Freericks2005}.
The importance of initial conditions was already demonstrated in nonequilibrium
DMFT studies of the time evolution of the FKM after a quench \cite{EcksteinPRL2008}.
We partially address this problem by using two different system potentials.

It is advantageous to rewrite the full model in a different basis
using the fact that the $f$-particle occupation numbers $f_{i}^{\dagger}f_{i}^{\phantom{\dagger}}$
commute with the entire Hamiltonian and are thus good quantum numbers.
This allows us to replace $f_{i}^{\dagger}f_{i}^{\phantom{\dagger}}$
by its eigenvalues $w_{i}=0$ or $w_{i}=1$ and write a partial Hamiltonian
for a particular classical configuration $w$. This significantly
simplifies the problem. The system Hamiltonian for a chosen configuration
$w$ reads 
\begin{equation}
H_{S}^{w}=\sum_{i,j}h_{ij}d_{i}^{\dagger}d_{j}^{\phantom{\dagger}}+\sum_{i}\varepsilon_{i}^{f}w_{i}=\sum_{\alpha}\tilde{\varepsilon}_{\alpha}\tilde{d}_{\alpha}^{\dagger}\tilde{d}_{\alpha}^{\phantom{\dagger}}+\sum_{i}\varepsilon_{i}^{f}w_{i},\label{eq:H1tr}
\end{equation}
where $h_{ij}=\left(Uw_{i}+\varepsilon_{i}\right)\delta_{ij}-t_{ij}$
and where we have diagonalized the first term by formally applying
a simple unitary transformation

\[
d_{i}^{\phantom{\dagger}}=\sum_{\alpha}\mathcal{U}_{i\alpha}\tilde{d}_{\alpha}^{\phantom{\dagger}},\;d_{j}^{\dagger}=\sum_{\beta}\tilde{d}_{\beta}^{\dagger}\mathcal{U}_{\beta i}^{\dagger},
\]
with $\tilde{\varepsilon}_{\alpha}$ being the eigenvalues of matrix
$\mathbf{h}$ and where matrix $\mathcal{U}$ consists of the related
eigenvectors arranged in columns. The matrix $\mathcal{U}$ can be
computed numerically for a finite system and it can be chosen to be
real. We can formally apply an equivalent transformation also to lead
terms 
\[
c_{l,n}^{\phantom{\dagger}}=\sum_{k}\mathcal{V}_{l,nk}\tilde{c}_{l,k}^{\phantom{\dagger}},\;c_{l,n}^{\dagger}=\sum_{k}\tilde{c}_{l,k}^{\dagger}\mathcal{V}_{l,kn}^{\dagger},
\]
leading to 
\begin{eqnarray}
H_{\mathrm{lead}}^{l} & = & \sum_{k}\tilde{\epsilon}_{l,k}\tilde{c}_{l,k}^{\dagger}\tilde{c}_{l,k}^{\phantom{\dagger}},\\
H_{\mathrm{hyb}}^{l,w} & = & \sum_{\alpha,k}\tilde{\gamma}_{l,k\alpha}\tilde{c}_{l,k}^{\dagger}\tilde{d}_{\alpha}^{\phantom{\dagger}}+\tilde{\gamma}_{l,\alpha k}^{\dagger}\tilde{d}_{\alpha}^{\dagger}\tilde{c}_{l,k},\label{eq:H2_3tr}
\end{eqnarray}
where $\tilde{\gamma}_{l,k\alpha}=-\gamma_{l}\sum_{\left\langle i,n\right\rangle }\mathcal{V}_{l,kn}^{\dagger}\mathcal{U}_{i\alpha}$
keeps track of the actual geometry of the system-lead interface and
therefore can not be taken constant.

The transformed Hamiltonian (\ref{eq:H1tr})-(\ref{eq:H2_3tr}) describes
a relatively simple noninteracting model where both the ``level''
energies $\tilde{\varepsilon}_{\alpha}$ and the hybridization strength
$\tilde{\gamma}_{l,\alpha k}$ are functions of the configuration
$w$. The nonequilibrium transport in this kind of models is a well
studied problem \cite{Jauho1994} and the exact form of the steady-state
Green's functions is given by\cite{Haug-Jauho} 
\begin{eqnarray}
\mathbf{G}^{r,a}\left(E\right) & = & \mathbf{g}^{r,a}\left(E\right)+\mathbf{g}^{r,a}\left(E\right)\mathbf{\Sigma}^{r,a}\left(E\right)\mathbf{G}^{r,a}\left(E\right),\label{eq:Gra}\\
\mathbf{G}^{<}\left(E\right) & = & \mathbf{G}^{r}\left(E\right)\mathbf{\Sigma}^{<}\left(E\right)\mathbf{G}^{a}\left(E\right).
\end{eqnarray}
Here, $\mathbf{G}^{r\,(a)}$ is the retarded (advanced) Green's function
of the whole structure, $\mathbf{G}^{<}$ is the lesser Green function
and $\mathbf{g}^{r\,(a)}\left(E\right)$ is the retarded
(advance) Green's function of the bare system with components 
\begin{equation}
g_{\alpha\beta}^{r,a}\left(E\right)=\frac{\delta_{\alpha\beta}}{E-\tilde{\varepsilon}_{\alpha}\pm i0}.
\end{equation}
The total tunneling self-energies $\mathbf{\Sigma}^{r,a,<}=\mathbf{\Sigma}_{L}^{r,a,<}+\mathbf{\Sigma}_{R}^{r,a,<}$
have the components 
\begin{eqnarray*}
\Sigma_{l,\alpha\beta}^{r,a,<}\left(E\right) & = & \sum_{k}\tilde{\gamma}_{l,\alpha k}^{*}\mathbf{g}_{l,k}^{r,a,<}\left(E\right)\tilde{\gamma}_{l,k\beta},
\end{eqnarray*}
where $\mathbf{g}_{l}^{r,a,<}\left(E\right)$ is the noninteracting
Green's function of the isolated lead $l$. To simplify the analysis,
we assume leads with zero hopping in the direction parallel to the
system-lead interface, i.e., the leads consist of identical
semi-infinite chains with hopping $t_{l}$. The exact self-energies
for this choice of leads can be found analytically \cite{RyndykBook,Cizek2004}
and read 

\begin{eqnarray}
\Sigma_{l,\alpha\beta}^{r,a}(E) & = & \Lambda_{l,\alpha\beta}(E)\mp\frac{i}{2}\Gamma_{l,\alpha\beta}(E),\nonumber \\
\Sigma_{l,\alpha\beta}^{<}(E) & = & i\Gamma_{l,\alpha\beta}(E)\,f_{l}(E-\mu_{l}),\nonumber \\
\Gamma_{l,\alpha\beta}(E) & = & 2\pi\gamma_{l}^{2}\mathcal{U}_{\alpha\beta}^{\{s^{l}\}}\rho_{l}(E),\label{eq:Gamma}\\
\Lambda_{l,\alpha\beta}(E) & = & \frac{2\gamma_{l}^{2}}{B^{2}}\mathcal{U}_{\alpha\beta}^{\{s^{l}\}}\left(E-\mu_{l}\right),\nonumber \\
\mathcal{U}_{\alpha\beta}^{\{s^{l}\}} & = & \sum_{\begin{array}{c}
i\in\{s^{l}\}\end{array}}\mathcal{U}_{\beta i}^{\dagger}\mathcal{U}_{i\alpha}^{\phantom{\dagger}},\nonumber 
\end{eqnarray}
where $\{s^{L,R}\}$ are the sets of system lattice positions at the
left and right interfaces, $f_{l}(E)$ is the Fermi function, and $\mu_{l=L,R}$
is the chemical potential of the leads. The leads are characterized
by a smooth surface density of states (DOS):

\[
\rho_{l}\left(E\right)=\frac{2}{\pi B^{2}}\sqrt{B^{2}-\left(E-\epsilon_{l}\right)^{2}},
\]
with band half width $B=2t_{l}$ centered around $\epsilon_{l}$. 

The tunneling current for fixed configuration $w$ has a compact Landauer-Büttiker
form,\cite{Haug-Jauho} 
\begin{equation}
I^{w}=\int\frac{dE}{2\pi}\,\mathscr{T}^{w}\left(E\right)\left[f(E-\mu_{L})-f(E-\mu_{R})\right],\label{eq:Current}
\end{equation}
with the transmission function 
\begin{equation}
\mathscr{T}^{w}\left(E\right)=\mathrm{Tr_{d}\,}\left[\mathbf{\Gamma}_{L}\left(E\right)\mathbf{G}^{r}\left(E\right)\mathbf{\Gamma}_{R}\left(E\right)\mathbf{G}^{a}\left(E\right)\right],\label{eq:Transmission}
\end{equation}
where the trace goes over the $d$-electron subsystem.
In the following we set $\mu_{L}=\epsilon_{L}$ and $\mu_{R}=\epsilon_{R}$,
which corresponds to half-filled lead bands and we introduce a finite
voltage drop as $V=\mu_{L}-\mu_{R}$ with condition $\mu_{L}=-\mu_{R}$.

It is worth noting that the numerical solution of Eq.~(\ref{eq:Gra}),
which represents a system of linear equations for complex matrices
of the size $L\times L$, together with the matrix multiplication
in Eq.~(\ref{eq:Transmission}), present the bottleneck for the numerical
evaluation. This is because, despite the broadening provided by the
self-energy, the Green's functions can still contain sharp $\delta$-function-like
features, especially at low temperatures and for strong interaction
$U$. The evaluation of the current therefore requires a dense mesh
of energies $E$ (see e.g. Fig.~\ref{fig:TransFlat}(h)). Fortunately,
this part can be easily parallelized.

So far, we have addressed only the problem for a fixed configuration
$w$. The calculation of average values of any operator $\hat{O}$
for the $d$ electrons requires a trace over the $f$ particles as
well. Considering the above stated initial condition and assuming
that the leads are not affected by the system, the trace over the
classical degrees of freedom reduces to a simple formula 
\begin{equation}
\left\langle \hat{O}\right\rangle =\frac{1}{Z}\sum_{w}e^{-\beta F\left(w\right)}\left\langle \hat{O}\right\rangle _{d},\label{eq:Aver}
\end{equation}
where 
\begin{equation}
F\left(w\right)=-\frac{1}{\beta}\sum_{\alpha}\ln\left[1+e^{-\beta\tilde{\varepsilon}_{\alpha}}\right]+\sum_{i}\varepsilon_{i}^{f}w_{i},\label{eq:Free}
\end{equation}
with $Z=\sum_{w}e^{-\beta F\left(w\right)}$ being the partition function
and $\left\langle .\right\rangle _{d}$ being the trace over the $d$-electron
subsystem for fixed $w$ \cite{MaskaPRB2006}. E.g., the total (bulk)
DOS of the isolated system is defined as $\mathrm{DOS}(E)=\mathrm{Tr}_{w}\sum_{\alpha}\delta\left(E-\varepsilon_{\alpha}\right)/L$
and the DOS of the coupled heterostructure is defined as $\mathrm{DOS}(E)=\mathrm{Tr}_{w}\mathrm{Tr}_{d}i\left[\mathbf{G}^{r}\left(E\right)-\mathbf{G}^{a}\left(E\right)\right]/2\pi L$,
where $\mathrm{Tr}_{w}\equiv\frac{1}{Z}\sum_{w}e^{-\beta F\left(w\right)}$
. Calculation of the averages of operators which reflect the real-space 
distributions requires us to retrace the unitary transformation.
For example, the average $d$-electron occupation on a particular
site $i$ of a coupled system is given by 
\begin{eqnarray}
\left\langle n_{d}^{i}\right\rangle  & = & -i\mathrm{Tr}_{w}\int dE\,\sum_{\alpha,\beta}\mathcal{U}_{i\alpha}\mathcal{U}_{\beta i}^{\dagger}G_{\alpha\beta}^{<}\left(E\right),\label{eq:nd}
\end{eqnarray}
and the local density of states (LDOS) can be calculated using 
\begin{eqnarray}
\mathrm{LDOS_{\mathit{i}}(E)} & = & \frac{i}{2\pi L}\mathrm{Tr}_{w}\left[\sum_{\alpha,\beta}\mathcal{U}_{i\alpha}\mathcal{U}_{\beta i}^{\dagger}G_{\alpha\beta}^{r}\left(E\right)\right.\label{eq:LDOS}\\
 &  & \left.-\sum_{\alpha,\beta}\mathcal{U}_{\alpha i}^{\dagger}\mathcal{U}_{i\beta}G_{\alpha\beta}^{a}\left(E\right)\right].\nonumber 
\end{eqnarray}
The power of this approach lies in the fact that the averaging in
Eq.~(\ref{eq:Aver}) can be performed by a simple sign-problem-free
Monte Carlo method \cite{MaskaPRB2006,ZondaSSC2009,Zonda2012,Huang2017}.
In this method, the classical configuration $w$ is updated following
the Metropolis algorithm, where the difference in the free energy,
Eq.~(\ref{eq:Free}), for different $w$ is used to build a Markov
chain. To obtain the free energy a single-particle quantum problem
is solved exactly in every Monte Carlo step by numerical diagonalization
which can be done efficiently \cite{Motome1999}. It is worth noting
that this method is not limited to the equilibrium or steady state
as the time evolution is accessible as well \cite{Herrmann2018}.

Because of the broadening of the Green's functions, provided by the
coupling to the leads, we can also calculate the zero-temperature
characteristics. In this case, the Monte Carlo averaging can be avoided.
All that is needed, is the correct $f$-electron ground-state configuration.
We calculate the ground states using a simple simulated annealing
method\cite{Cerny1985}, which is similar to the Monte Carlo method
presented here, with the difference that the ground-state energy $E_{GS}(w)=\sum_{\alpha=1}^{N_{d}}\tilde{\varepsilon}_{\alpha}+\sum_{i}\varepsilon_{i}^{f}w_{i}$
for a particular configuration is used instead of $F(w)$ to calculate
the weights in the annealing process.

\section{Results\label{sec:Results}}

Because of the vastness of the parameter space we have restricted
the present paper to symmetric couplings of $\gamma=\gamma_{L}=\gamma_{R}=2t$
and $4t$. These values provide a sufficient broadening of
the transmission function even for zero temperature ($T=0$). We also
use relatively broad DOSs for the leads with the half width $B=B_{L}=B_{R}=10t$
or $20t$, which are, however, still narrow enough to study the
effects related to a finite band.

As already stated, the profiles of $\varepsilon^{d}$ and $\varepsilon^{f}$
can be influenced by the leads and should be, in general, calculated
self-consistently. Following Ref.~\onlinecite{Okamoto2008} we instead
consider two limiting cases: a flat potential, where $\varepsilon_{i}^{\mathrm{flat}}=\varepsilon_{i}^{d}=\varepsilon_{i}^{f}=-U/2$
(Fig.~\ref{fig:Schema}(b)) for every lattice point $i$; and a tilted
potential $\varepsilon_{i}^{\mathrm{tilt}}=\varepsilon_{i}^{d}=\varepsilon_{i}^{f}=-U/2+n_{x}(\epsilon_{L}-\epsilon_{R})/(L_{x}+1)$
(Fig.~\ref{fig:Schema}(c)), where $n_{x}$ numbers the layers from
left to right. Note, that for the purpose of this paper we use the
term ``layer'' for a single chain parallel to the interface ($y$ direction).
Both potential profiles are centered around $-U/2$. This choice fixes
the half-filling condition at zero chemical potential for the whole
system but not in every particular layer. We assume that the realistic
profile should be in between these two cases. These two choices can
be also understood as two different initial conditions. This follows
from the fact that the $f$ electrons are integrals of motion. Their
distribution is fixed by the initial thermalization. The flat potential
describes a situation where the system was initially completely isolated
from the leads and for the tilted potential the voltage drop heavily
influenced the system potential before the coupling was switched on.

\subsection{Flat potential }

We start our analyses with the flat electrostatic potential, which
we suppose is more realistic than the tilted one, at least for a large
system. Our primary goal is to address two rather general questions, namely, 
how the coupling to the leads and the finite voltage influence
the phases of the FKM and how these phases, if still present, influence
the charge transport in the heterostructure. Therefore we first discuss
the phase diagram of the spinless FKM.

\subsubsection{Phase diagram and equilibrium properties\label{subsec:PhDF}}

\begin{figure}
\includegraphics[width=1\columnwidth]{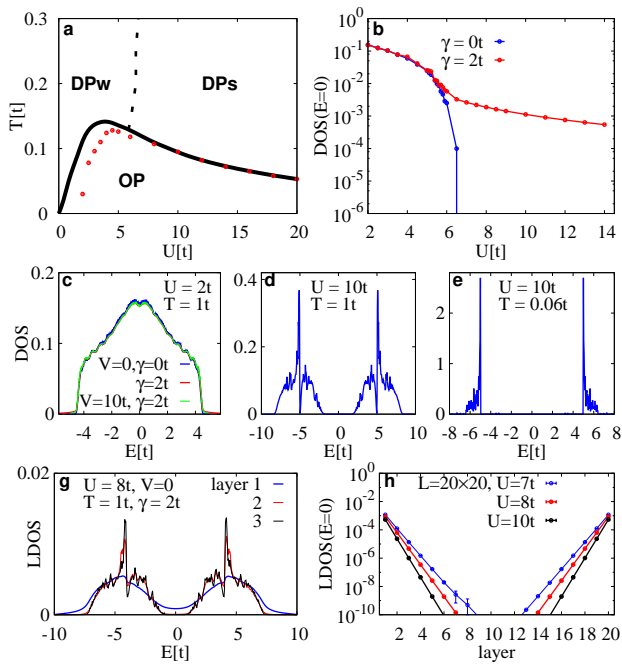}\caption{(a) Simplified phase diagram of the isolated system with ordered phase
(OP) and disordered phases in weak (DPw) and strong (DPs) interaction
regimes. The lines represent phase boundaries for the flat potential
$\varepsilon^{\mathrm{flat}}$. The dots show an estimate for $\varepsilon^{\mathrm{tilt}}$
at $V=5t$, taken from the position of the maxima of the approximated
specific heat \cite{MaskaPRB2006} for $L=20\times20$. (b) Equilibrium
$d$-electron DOS at $E=0$ for coupled (red) and decoupled (blue)
systems calculated in the vicinity of DPw-DPs transition for the flat
system potential for $T=1t$ and plotted as a function of $U$. (c-e)
Illustration of the typical $d$-electron DOS for DPw (c), DPs (d), and
OP (e) from the phase diagram (a). Blue lines have been calculated
for the isolated system, and red and green lines in (c) have been 
calculated for the coupled
system. (g) Local DOS of the coupled system at equilibrium for the
first three system layers parallel to the interface. (h) Local DOS
of the coupled system calculated at the Fermi level for all system
layers and various values of $U$ in the Mott regime. In all cases
the system size is $L=20\times20$.\label{fig:Phd}}
\end{figure}

The general phase diagram of the spinless FKM is profoundly rich and
includes stable exotic orderings such as stripes and various charge
segregations \cite{Lemanski1995,LemanskiPRL2002,Cencarikova2011,Zonda2012}.
Moreover, the phase diagram is fairly complicated even at half filling,
where the FKM with flat potential is at low temperature always in
the CDW phase for any bipartite lattice \cite{Brandt1989,Brandt1990,Brandt1991,ChenPRB2003,ZondaSSC2009}.
In Fig.~\ref{fig:Phd}(a), a simplified version of the phase diagram
is plotted for an isolated 2D FKM system. The three main regions that
are crucial for our study of transport are the low-temperature ordered
phase (OP), the disordered phase for weak interaction (DPw), and the
disordered phase for strong interaction (DPs). The equilibrium $d$-electron
DOS in OP (Fig.~\ref{fig:Phd}(e)) contains a CDW gap with a width
equal to $U$ at zero temperature. The width of the CDW gap does not
change with increasing temperature, but rather the gap is filled in
by subgap states \cite{Hassan2007,MatveevPRB2008,Lemanski2014}. The
gap is completely closed in DPw (Fig.~\ref{fig:Phd}(c)) and a narrower
Mott gap stays open between Hubbard-like bands in DPs (Fig.~\ref{fig:Phd}(d)).
Nevertheless, a CDW pseudo gap can be present in the DOS even at high
temperatures. 

Because of the absence of the gap in the DOS (Fig.~\ref{fig:Phd}(c)), the DPw was initially considered to be metallic in two dimensions \cite{MaskaPRB2006,ZondaSSC2009}, but recent studies on bigger clusters showed that in the thermodynamic limit it is actually an Anderson insulator phase  and that for any finite system the DPw also contains a crossover from the Anderson insulator to a broad weak localization phase at weak interactions \cite{Antipov2016,Lee81}.

It is worth noting that the phase diagram is in principle even more
complicated. For example, for some parameters it is possible to have
a CDW gap in the DOS but a nonzero DOS at the Fermi level \cite{Hassan2007,Lemanski2014,Matveev2016}.
Here we focus on the phases mentioned above.

We first investigate the influence of the leads in equilibrium. The
phase boundary between OP and disordered phases is not affected by
finite $\gamma$. This is because the phase transition is mostly driven
by the ordering of $f$ particles and, as their occupations are integrals
of motion, the full system always keeps the thermalized $f$-particle
distribution of the decoupled system. This is, however, not necessarily
true for the DPs-DPw phase boundary, which mostly reflects the transition
from finite to zero gap in the DOS of $d-$electrons.

We show a comparison of the total DOS at the Fermi level ($E=0$)
for the isolated and coupled system (with $\gamma=2t$) in Fig.~\ref{fig:Phd}(b).
It is evident, that the coupling closes the Mott-like gap otherwise
clearly developed in the DPs. The reason is the leaking of the leads
metallic density of states into the system. This is illustrated in
Fig.~\ref{fig:Phd}(g), where we plot the LDOS calculated for the
first three system layers parallel to the interfaces. The LDOS of
the first layer (blue line) is strongly broadened by the coupling.
The effect quickly vanishes with increasing distance from the interface.
The decay is exponential for strong interaction, which is in compliance
with DMFT studies \cite{Freericks2004,FreericksBook2006}. We show
this in Fig.~\ref{fig:Phd}(g), where we plot the dependence of LDOS$(E=0)$
on the distance from the left interface for $L=20\times20$ and three
values of $U$. The leaking of the leads' DOS into the system is the
dominant contribution to the LDOS$(E=0)$ for strong interaction (Fig.~\ref{fig:Phd}(g)),
because there the Mott-like gap is clearly developed for the isolated
system even for finite lattice sizes. Therefore, the exponential decay
of the LDOS$(E=0)$ with the distance from the interfaces reflects
directly the vanishing influence of the leads. Note, that the decay
clearly depends on $U$ and it is therefore more complicated for intermediate
and weak coupling where, in addition, the gap is not yet opened even
for a decoupled system. It is therefore challenging to analyze quantitatively
the effect of the leads on DPs-DPw transition. Nevertheless, we will
readdress the question in the next section, where we discuss the transport
properties. 

\begin{figure}
\includegraphics[width=1\columnwidth]{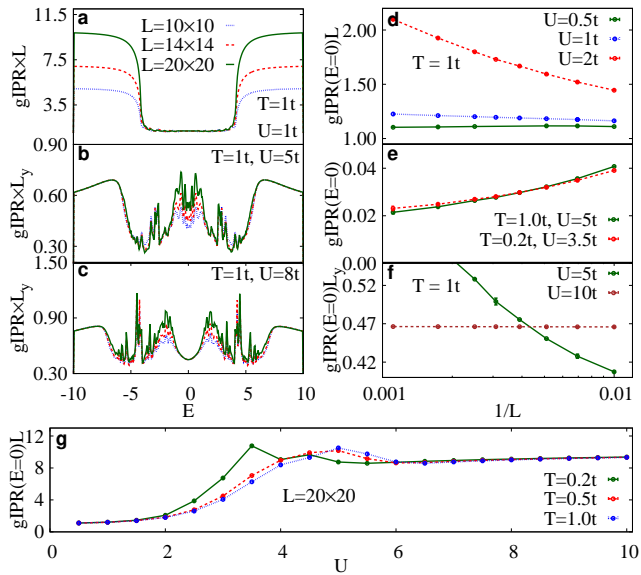}\caption{(a-c) The generalized inverse participation ratio calculated for the coupled
system and for different $U$ and $L$. The results plotted in (a)
are multiplied by $L$ and in (b,c) are multiplied by $L_{y}.$ (d-f)
The size scaling of the generalized inverse participation ratio at
the Fermi level in three regimes discussed in the main text. (g) The
generalized inverse participation as a function of $U$ for different
temperatures and $L=20\times20$. All figures show equilibrium state
($V=0$) with coupling $\gamma=2t$ and half bandwidth of the bands
set to $B=10t$. \label{fig:gIPR}}
\end{figure}

The broadening of the DOS of the coupled system allows us to define
the averaged generalized inverse participation ratio (gIPR) \cite{Murphy2011,Perera2018}
\[
\mathrm{gIPR}(E)=\mathrm{Tr}_{w}\frac{\sum_{i}\mathrm{LDOS_{\mathit{i}}^{2}\mathit{(E,w)}}}{\mathrm{DOS}^{2}(E,w)},
\]
 for the coupled system without the necessity to regularize the possible
$\delta$-functions. The inverse participation ratio and its generalization
are used for the identification of localization in isolated strongly
correlated electron systems in equilibrium \cite{Evers2008,Murphy2011,Perera2018}.
Here we show that it can be useful also for the analysis of an open
system.

The $\mathrm{gIPR}$ scales as $1/L$ for completely itinerant system
states. Considering the exponential decay of the influence of the
lead DOS shown in Fig.~\ref{fig:Phd}(g), the ratio scales as $\sim1/L_{y}$
for energies within the gap as here the dominant contribution comes
from the interfaces. The ratio should converge to a finite value with
increasing $L$ for localized system states and should be constant
for a sufficiently large system and strong localization. The inverse
participation ratio is therefore ideal for studying the complicated
transition from the Fermi gas at $U=0$ to the Mott-like phase for
$U\gtrsim6t$ \cite{Antipov2016}. 

Fig.~\ref{fig:gIPR} shows an analysis of the $\mathrm{gIPR}$ for
the coupled system in equilibrium. Subplots (a-c) depict its energy
dependence for weak ($U=1t$) intermediate ($U=5t$) and strong interaction
($U=8t$) and different lattice sizes. For weak interaction the $\mathrm{gIPR}$
scales as $\sim1/L$ (note that the data in Fig.~\ref{fig:gIPR}(a)
are multiplied by $L$) for a broad range of energies around zero
and significantly deviates from this scaling only for energies outside
the range of the isolated-system DOS where the main contributions
come from the leads' broadening. This is shown also in Fig.~\ref{fig:gIPR}(d),
where we plot the scaling of the $\mathrm{gIPR}$ at the Fermi level
for small $U$. The value of $\mathrm{gIPR}(E=0)\times L$ is practically
constant for $U\lesssim t$ and one can conclude that in this region
the states are predominantly delocalized for a finite number of layers.

For strong interaction $U$ (Fig.~\ref{fig:gIPR}(c,f)), $\mathrm{gIPR}$
scales as $1/L_{y}$ in the region of the Mott gap, as well as for
energies outside the full width of the DOS of the isolated system,
as expected for a gapped system. The scaling in the intermediate interaction
region ($3t\lesssim U\apprle6t$) is not that straightforward. The
dependence of $\mathrm{gIPR}(E=0)$ on $U$ shows a maximum in this
region evident in Fig.~\ref{fig:gIPR}(g) and the scaling (Fig.~\ref{fig:gIPR}(e))
seems to point to a finite value of $\mathrm{gIPR}(E=0)$ in the thermodynamic
limit. Both these results indicate an Anderson-like localization for
states near $E=0$ which is in compliance with a previous study of
the isolated system by Antipov {\it et al.}, \cite{Antipov2016}. However,
a more thorough analysis on a significantly larger system is necessary
to confirm this conjecture. 

Note, that the change of the temperature does not play a significant
role if $T$ is much higher than the critical temperature $T_{c}(U)$
of the order-disorder transition (Fig.~\ref{fig:gIPR}(g)). However,
this changes when the temperature approaches $T_{c}$. The sensitivity
of $\mathrm{gIPR}(E=0)$ on temperature is most notable in the intermediate
interaction regime. Nevertheless, we can conclude, that all main phases
of the FKM are still present even for the coupled heterostructure
which includes different localization regimes in the DPw.

\subsubsection{Transport properties \label{subsec:TransportA}}

We start our discussion of the transport properties for the flat potential
by considering a finite $L=20\times20$ system and first explain its
most important features. Only afterwards we discuss the effects of
the system size on the transport in detail (Sec.~\ref{subsec:SSET}).

\begin{figure}
 \includegraphics[width=1\columnwidth]{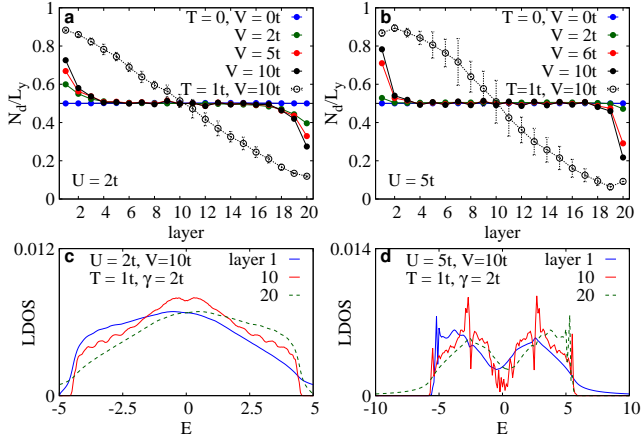}\caption{(a,b) Normalized $d$-particle occupations per layer parallel to the
system-lead interfaces for a coupled system with $\gamma=2t$, $B=10t$,
flat system potential, with interactions (a) $U=2t$, and (b) $U=5t$. The
filled circles show the zero temperature results for various $V$.
The open circles represent the high-temperature result for $T=1t$
and $V=10t$. (c,d) The nonequilibrium LDOS calculated at the interfaces
and central part of the system for high temperature ($T=1t$) and
high voltage $V=10t$. \label{fig:NEnd}}
\end{figure}

The real-space distribution of the $d$ electrons of the decoupled
system either is homogeneous or forms a CDW pattern. In both cases
the flat potential fixes the average $d$-electron occupancy to $L_{y}/2$
for every layer. The nonequilibrium distribution of the coupled system
shown in Fig.~\ref{fig:NEnd} is somewhat different. Following the
respective chemical potential of the coupled leads, the distribution
is elevated or lowered close to the interface for $T=0$. In the central
region, the distribution approaches $L_{y}/2$. At high temperatures,
the normalized occupations for very high voltage have nearly a linear
slope (see the open black circles in Fig.~\ref{fig:NEnd}). A significant
spatial difference at high voltage can be seen also in the LDOS shown
in Fig.~\ref{fig:NEnd}(c,d). The shifted leads' surface densities of
states $\rho_{L}(E)$ and $\rho_{R}(E)$ broaden the system LDOS asymmetrically.
As expected, the effect is most notable at the interfaces. For example,
the part of the system LDOS located outside the width of $\rho_{l}(E)$
of the neighboring lead is broadened only by the temperature. 

\begin{figure}
\includegraphics[width=1\columnwidth]{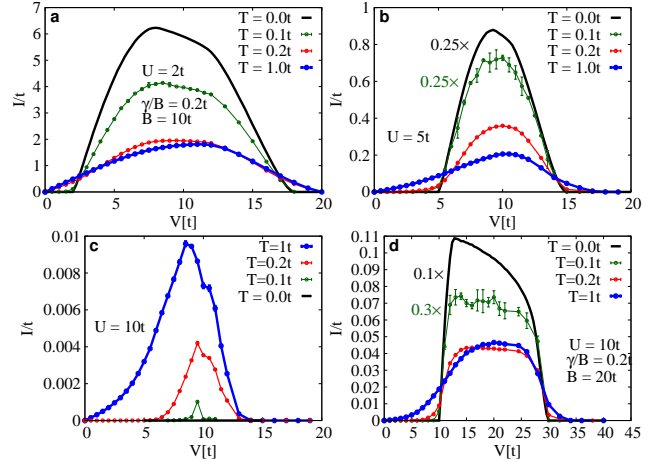}

\caption{Current-voltage characteristics for flat system potential, system
size $L=20\times20$, coupling $\gamma/B=0.2$ and parameters (a)
$U=2t$, $B=10t$; (b) $U=5t$, $B=10t$; (c) $U=10t$, $B=10t$; and
(d) $U=10t$, $B=20t$ . Note that the curves for the two lowest temperatures
in (b) are scaled by factors $0.25$ and in (d) by factors $0.1$
and $0.3$ to improve the visibility. \label{fig:CurrentFlat}}
\end{figure}

\begin{figure}
\includegraphics[width=1\columnwidth]{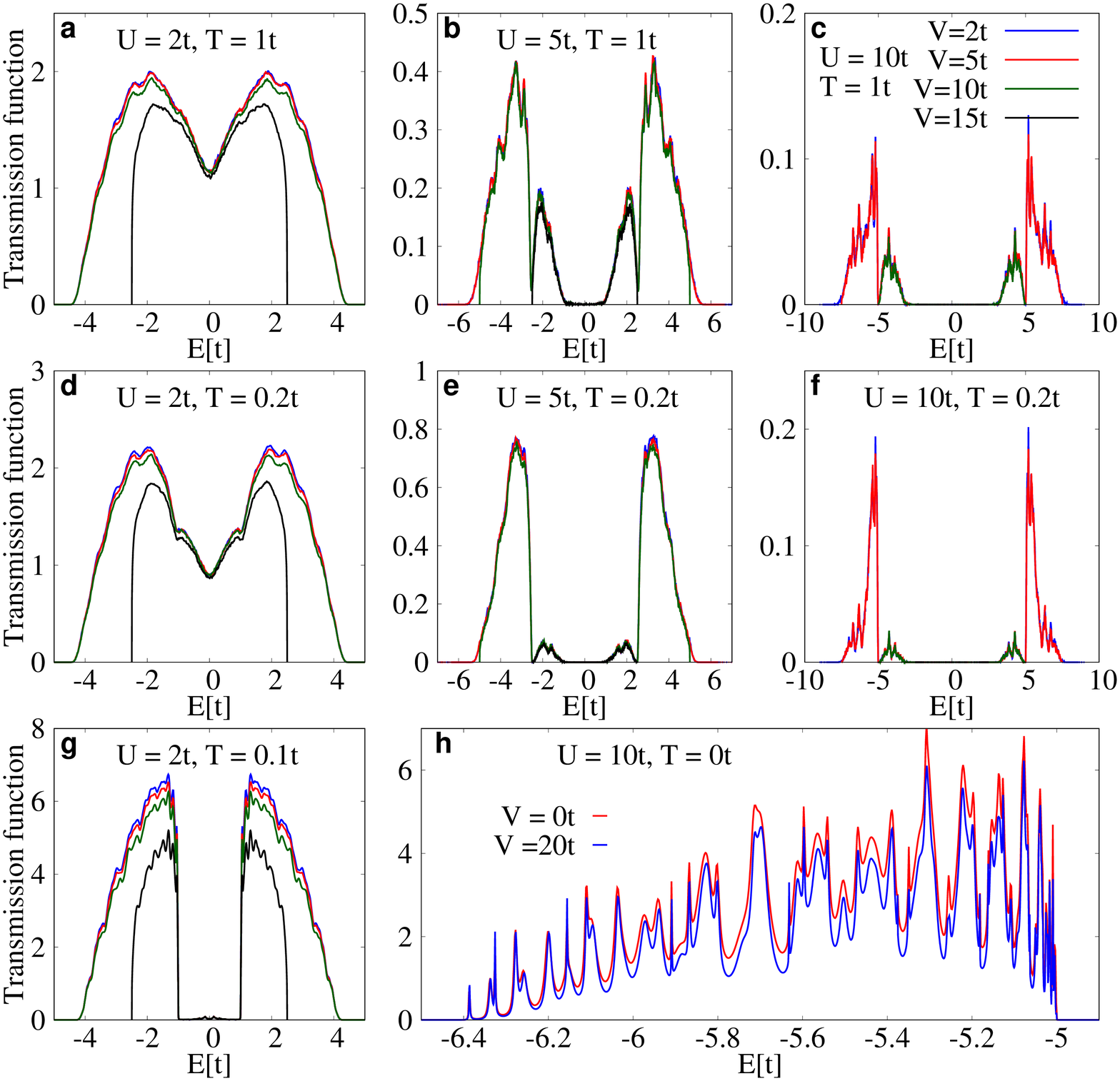}\caption{(a-g) Transmission function for the flat system potential for coupling
$\gamma=2t$, band half-width $B=10t$ and different values of $U$,
$T$, and $V$. (h) Detail of the Hubbard-like satellite for $\gamma=4t$,
band half width $B=20t$ and strong interaction $U=10t$ plotted for
the $V=0t$ and $20t$ cases. \label{fig:TransFlat}}
\end{figure}

Fig.~\ref{fig:CurrentFlat} shows examples of $I-V$ characteristics
of the FKM system for three values of $U$ which, according to Fig.~\ref{fig:Phd}(a),
represent the weak-interaction ($U=2t$), intermediate-interaction ($U=5t$) and 
strong-interaction
($U=10t$) case. The results in Fig.~\ref{fig:CurrentFlat}(a-c)
have been calculated using the coupling $\gamma=2t$ and a bandwidth
of $B=10t$. The current in Fig.~\ref{fig:CurrentFlat}(d) was obtained
for the same ratio $\gamma/B$ but a broader band, $B=20t$. The two
high-temperature cases $T=1t$ and $T=0.2t$ illustrate how the decreasing
temperature influences the transport in the disordered phases. Temperature
$T=0.1t$ addresses the transport just below the phase boundary and
$T=0t$ shows the system without thermal excitations. To understand
the $I-V$ characteristics, we also present in Fig.~\ref{fig:TransFlat}
the related transmission functions for selected voltages. The transmission
functions are particularly useful as they contain the most detailed
information about elastic transport. The nonequilibrium transmission
for small $V$ resembles the equilibrium DOS. Similarly to the total
DOS, and despite the strong modulation of the LDOS (Fig.~\ref{fig:NEnd}(c,d)),
the transmission function changes significantly with voltage only
when $V\gtrsim B$ (see, e.g., case $V=15t$ in Fig.~\ref{fig:TransFlat}(a,d,g)).
However, the most important features considering transport are much
more pronounced in the transmission function than in the DOS or LDOS
already in equilibrium. This is particularly true for the pseudo-gap
around $E=0$, which is evident even for high temperature and weak
interaction (compare Fig.~\ref{fig:NEnd}(c) with Fig.~\ref{fig:TransFlat}(a)).
A similar statement is true also for the typical sub-gap structures
for intermediate and strong interaction (Fig.~\ref{fig:TransFlat}(b,c)).
This suggests that the $f$-electron configurations responsible for
the closing of the CDW gap at high temperatures have low transmission
which can be again attributed to a significant localization of the
$d$ electrons. 

Most of the features of the current can be understood by following
the two windows confining the integration over the relevant part of
the transmission function in Eq.~(\ref{eq:Current}). The first one
is the Fermi function window, the width of which is proportional (and at zero
temperature identical) to $V$. The second one is the band window
$W=2B-V$ (see Fig.~\ref{fig:Schema}(b)), which restricts the width
of the transmission function (Eq.~(\ref{eq:Gamma})).

The basic profile of the current for $T=1t$ is quite similar for
all values of $U$. The current increases up to $V\sim B$ and then
decreases to zero. The increase of the current reflects the broadening
of the Fermi function window. Around $V=B$ the two windows change
their roles because for $V>B$ it holds that $W<V$ and the decrease of
the current reflects the decrease of $W$. This is a typical effect
of the finite bandwidth of the leads, but their shape and the width
of the DOS of the isolated system play a role as well. 

As expected, the maximal current decreases with increasing $U$ because
the $d$ electrons become more localized. However, whereas the current
maximum is increasing with decreasing temperature for $U=2t$ and
$5t$, it seems to rapidly vanish for $U=10t$ in Fig.~\ref{fig:CurrentFlat}(c).
This is again a direct consequence of the finite bandwidth of the
leads. For $U=10t$, $V>10t$, and $B=10t$ and regardless of the broad
Fermi window, the transmission function fenced by $W$ covers only
the sub-CDW gap structures in Fig.~\ref{fig:TransFlat}(c,f), which
vanish with decreasing $T$. Therefore, the current drops rapidly
with decreasing $T$ and also for $V>B$. To analyze the influence
of the Hubbard-like satellites on the current in the strong-interaction
regime, we have calculated the current also for broader bands $B=20t$
while keeping the ratio $\gamma/B=0.2$ (see Fig.~\ref{fig:CurrentFlat}(d)).
The results confirm the trend of increasing the current maximum with
decreasing temperature.

An opposite trend can be observed in the region where $V<U$ (and
$V\sim2B$). For $U=2t$, $T=1t$ and small voltages, the current
depends approximately linearly on $V$ but decreases with both increasing
$U$ and decreasing $T$. The decrease is a consequence of the opening
and deepening of the CDW pseudo gap with decreasing temperature and
opening of the Mott pseudo gap with increasing $U$ in the transmission
function. This can be seen in Fig.~\ref{fig:TransFlat}(g) where
the transmission function is almost negligible in the interval $\left|E\right|<U/2$
already for $T=0.1t$ and even at weak interaction. The increase of
the Fermi window to $V\sim U$ leads to a sharp increase of the current
in Fig.~\ref{fig:CurrentFlat} as now the window covers the states
outside the CDW gap as well.

\begin{figure}
\includegraphics[width=1\columnwidth]{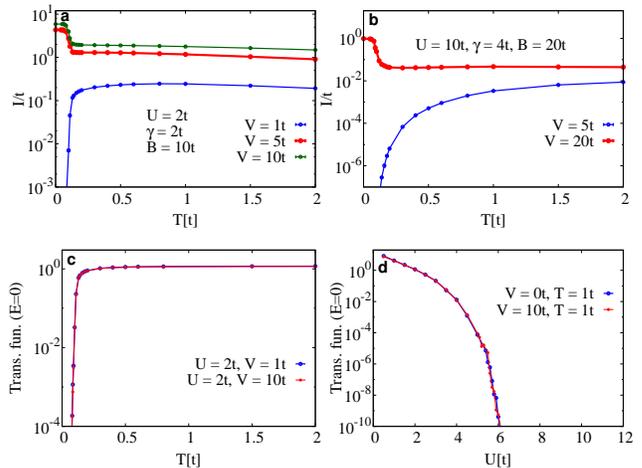}\caption{(a-b)Temperature dependence of the current for voltages below and above
the respective width of the CDW gap. (c) Transmission function $E=0$ as a function of temperature for $U=2t$ and $V=1t,\,10t$.
(d) Transmission function at $E=0$ as a function of $U$ for $T=1t$
and $V=0,\,10t$. \label{fig:FlatTdep}}
\end{figure}

We illustrate the different trends in different voltage regions again
in Fig.~\ref{fig:FlatTdep}(a,b), where we show the current as a
function of the temperature for voltages $V\lesssim U$ and $V\sim B$
(close to optimal voltage with maximal current). The opposite trends
are clearly demonstrated for temperatures near and below the DP-OP
transition. For $V\lesssim U$, the current drops around the critical
temperature. This change is steplike for weak interaction (where
the high-temperature phase is DPw) and gradual on the logarithmic
scale otherwise. On the other hand, the current increases dramatically
just below the critical temperature for $V\sim B$. Interestingly,
this means that the current can signal the order-disorder phase transition
of the system differently for low and optimal voltage.

This result can be understood by focusing on three tendencies. First,
the Fermi window becomes sharper with decreasing temperature, which
puts more ``weight'' on the transport through the central regions
of the DOS.  The second tendency is that the Hubbard-like maxima
in the transmission function are becoming narrower and higher with
decreasing temperature (compare Fig.~\ref{fig:TransFlat}(c) and
(h)). The sharp Fermi window, therefore, ensures a sharp increase
of the current in the interval of voltage that covers the Hubbard-like
satellites. This is most profoundly seen in Fig.~\ref{fig:CurrentFlat}(d).
For zero temperature, the current sharply increases in the range $U<V\lesssim13t$,
which covers exactly the Hubbard-like satellites in the transmission
function (see the detail in Fig.~\ref{fig:TransFlat}(h)). For higher
voltages, the current decreases. This is initially caused by a small
decrease of the transmission function shown in Fig.~\ref{fig:TransFlat}(h)
but mostly by closing of the band window $W$. The third tendency
is the rapid disappearing of the CDW in-gap states with decreasing
temperature. For small voltages ($V<U$) and low temperatures, this
leads to a drop in the current. The difference in the temperature
dependencies between the weak and strong intersection reflects the
fact that for strong interaction there is already a Mott gap in the
DOS even above the critical temperature.

We can therefore conclude that the three typical phases of the 2D
FKM have a qualitatively different influence on the charge transport
properties reflected in current characteristics. Moreover, the transition
between the phases can be seen directly from the transmission function.
We illustrate this in Fig.~\ref{fig:FlatTdep}(c,d) where the transmission
function at $E=0$ is plotted as a function of temperature (c) and
$U$ (d). Note, that at equilibrium the transmission function
is qualitatively equivalent to the conductance and it is relatively
insensitive to voltage at $E=0$ (see the overlap of the curves in
Fig.~\ref{fig:FlatTdep}(c,d)). Therefore, the sudden drop of the
transmission over several magnitudes that signals the critical temperature
in Fig.~\ref{fig:FlatTdep}(c) points to a qualitative change in
the character of transport between DPw and OP. Even more interesting
is the exponential drop of the transmission with increasing $U$ in
Fig.~\ref{fig:FlatTdep}(d). Although the total nonequilibrium DOS
plotted in Fig.~\ref{fig:Phd}(b) shows that the Mott gap in DPs
is closed by coupling to the leads, the drop of the transmission function
points to a similar critical $U$ as the DOS of the decoupled system
even after introducing a relatively high voltage ($V=10t$). This
reflects the rapid decrease of the influence of the leads on the LDOS
(Fig.~\ref{fig:Phd}(g,h)) with the distance from the interface.
Consequently, the DPw-DPs phase boundary seems to be unaffected by
the coupling to the leads and clearly manifests itself in the transport
properties.

\subsubsection{System size effects\label{subsec:SSEF} }

\begin{figure}
\includegraphics[width=1\columnwidth]{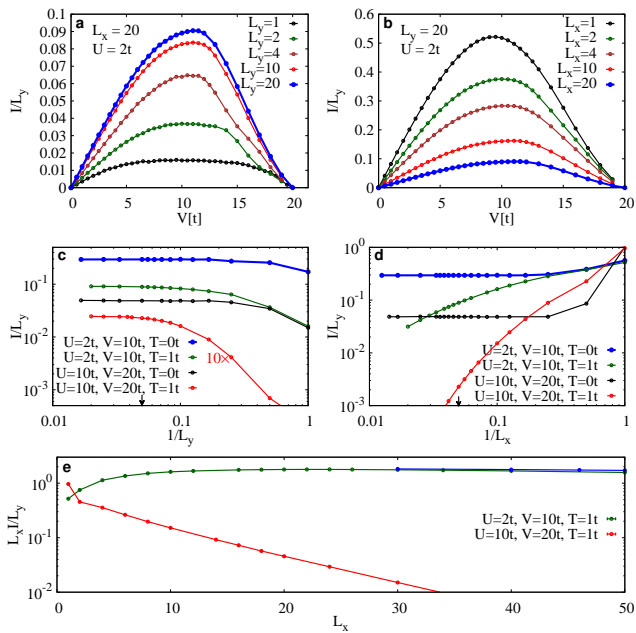}\caption{Current densities plotted: (a) as a function of voltage for a system
with fixed $L_{x}=20$ and different $L_{y}$ and $T=1t$; (b) as a function
of voltage for a system with fixed $L_{y}=20$ calculated for different
numbers of layers $L_{x}$ and $T=1t$; (c) as a function of inverse linear
size $1/L_{y}$ for $L_{x}=20$, $\gamma=0.2B$ , $U=2t$, $U=10t$,
optimal $V$, and zero temperature, where the current density for $U=10t$,
$T=1t$ was multiplied by $10$ for the sake of clarity; (d) as a function
of inverse linear size $1/L_{x}$ for a system with fixed $L_{y}=20$,
$\gamma=0.2B$, $U=2t$, $U=10t$ and optimal $V$, for high and zero
temperature, and (e) for a system with $L_{y}=20$ (green and red) and
$L_{y}=30$ (blue), where the current densities are multiplied by
$L_{x}$. \label{fig:FinSizFlat}}
\end{figure}

In this section, we consider the effect of the system size on the
transport properties. In Fig.~\ref{fig:FinSizFlat}(a), the current
densities ($I/L_{y}$) are plotted as functions of the voltage for
high temperature ($T=1t$), flat potential, weak coupling $U=2t$,
and different $L_{y}$, starting with a single chain ($L_{y}=1$).
The current density rapidly saturates with increasing $L_{y}$. This
is a general feature as we show in Fig.~\ref{fig:FinSizFlat}(c),
where the current densities are plotted for weak ($U=2t$) and strong
interaction ($U=10t$) at zero and high temperature as functions of
$1/L_{y}$. The chosen voltages are close to their optimal values,
where the biggest differences in current density for various $L_{y}$
are observed. All four curves are practically saturated at $L_{y}=20$.
This means that the results obtained for $L_{y}=20$ are actually
a good representation of a system in the limit $L_{y}\rightarrow\infty$.

We next focus on the other dimension, $L_{x}$. Fig.~\ref{fig:FinSizFlat}(d)
shows examples of the current densities for fixed $L_{y}=20$ but
different $L_{x}$ starting with a single layer. For zero temperature,
the current density rapidly saturates with increasing $L_{x}$. This
is because for the flat potential, the localized $f$ particles at
zero temperature form a periodic (checkerboard) potential and there
are no further sources of scattering. We conclude that the system
$L=20\times20$ is sufficient for addressing transport through an infinite
zero-temperature 2D FKM system at half filling.

The situation is different for $T=1t$ where the current density decreases
with the increasing $L_{x}$ (Fig.~\ref{fig:FinSizFlat}(d)). Nevertheless,
the size dependence is different for weak and strong interaction $U$.
In (Fig.~\ref{fig:FinSizFlat}(e)), we plot the same current densities
multiplied by $L_{x}$ and as a function of $L_{x}$. Whereas for
strong interaction the current density drops much faster than linearly,
for $U=2t$ the scaled current density initially increases and then
seems to saturate with $L_{x}$. Note, that there is actually a small
decrease of $IL_{x}/L_{y}$ for $L_{x}>30$; however, this is related
to fixed value of $L_{y}$. We confirmed the saturation by studying
the dependence for $L_{y}=30$. Interestingly, this linear dependence
of the current density on the system size together with almost linear
$I-V$ characteristics for $V\lesssim5t$ in DPw points to a metal-like
behavior. This might come as a surprise as it was recently shown that
2D FKM is an insulator in all its phases \cite{Antipov2016}, but
it is actually consistent with the analysis of the $\mathrm{gIPR}$
for both isolated \cite{Antipov2016} and coupled system (Fig.~\ref{fig:gIPR})
which show a strong delocalization of the density of states for finite
systems at weak interactions. 

\begin{figure}
\includegraphics[width=1\columnwidth]{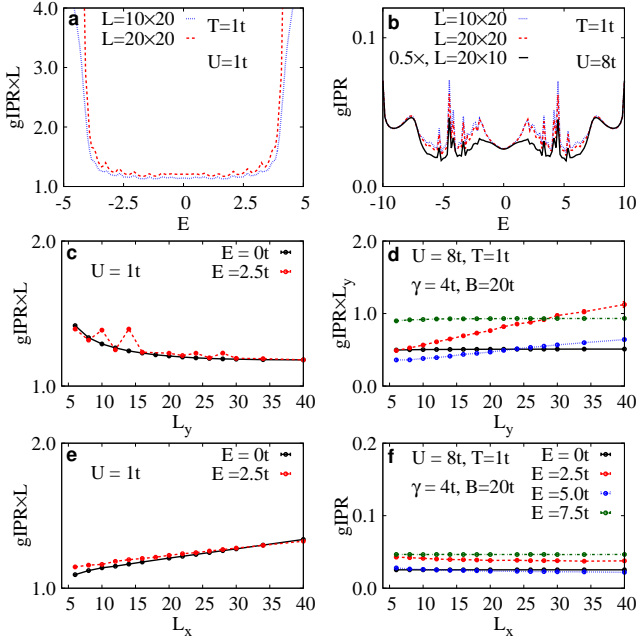}\caption{(a,b) The generalized inverse participation ratio for weak ($U=1t$)
and strong interaction ($U=8t$) and large voltage drop. (c,d) The
linear size scaling of the $\mathrm{gIPR}$ on $L_{y}$ for fixed
$L_{x}=20$ and the densities at various energies. (e,f) The linear
size scaling of the $\mathrm{gIPR}$ on $L_{x}$ for fixed $L_{y}=20$.
The weak-interaction limit is plotted for $V=10t$ , $\gamma=2t$,
and $B=10t$. The strong-interaction case is shown for $V=20t$ ,
$\gamma=4t$, and $B=20t$. \label{fig:gIPRV10}}
\end{figure}

We analyze this for the case of high voltage by studying again the
gIPR. Although the mutual shift of the $\rho_{L}$ and $\rho_{R}$
also introduces a spatial asymmetry in the LDOS, qualitative tendencies
are still obvious. Figs.~\ref{fig:gIPRV10}(a,b) show the $\mathrm{gIPR}$
profile for $E$ relevant for the transport at $V=10t$ for weak interaction
and $V=20t$ at strong interaction. Figs.~\ref{fig:gIPRV10}(c-f)
show its finite-size scaling for chosen energies. The wide plateau
in the energy profile of the weak coupling which scales roughly with
$1/L$ points to a prevailing delocalization. However, the scaling
is not as convincing as for the equilibrium case. This may point to
a stronger localization or could be a consequence of the discussed
shift of the LDOS. The scaling is much more clear for the strong interaction
($U=8t$). Figs.~\ref{fig:gIPRV10}(b,d) show that $\mathrm{gIPR}$
scales with $L_{y}$ only within the Mott gap and outside the normal
width of the DOS of the isolated system, where the broadening of the edge
layers dominates. The $\mathrm{gIPR}$ is practically constant for
all energies plotted in Fig.~\ref{fig:gIPRV10}(f) including the
energies within the isolated system DOS range. This points to a strong
localization of all states relevant for the transport which explains
the exponential drop of the current with the system width in Fig.~\ref{fig:FinSizFlat}(e).

\subsection{Tilted potential}

\subsubsection{Phase diagrams\label{subsec:PhDT}}

\begin{figure}
\includegraphics[width=1\columnwidth]{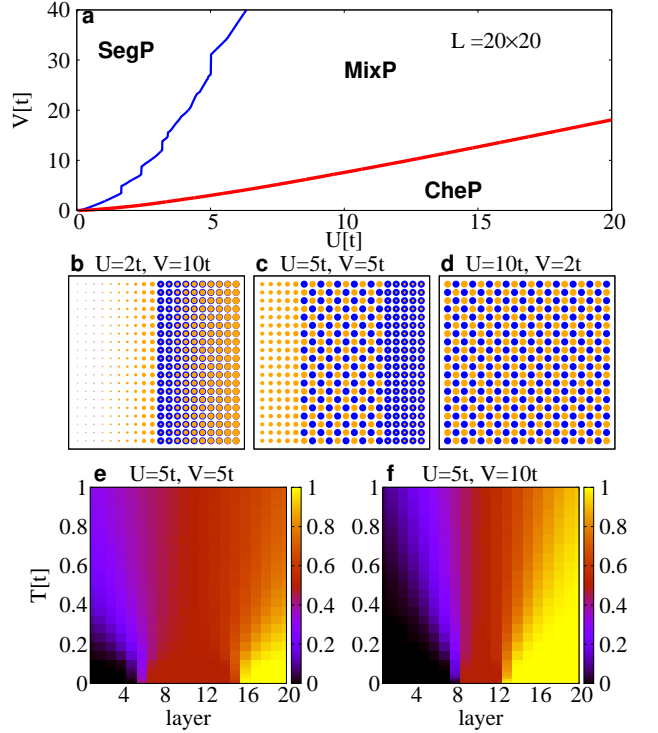}

\caption{(a) Ground-state phase diagram for tilted potential and $L=20\times20$
. Note, that the boundary between CheP and MixP is stable, but the
boundary between MixP and SegP depends on the system size. (b-d) The
representative ground-state configurations of $f$ particles (blue
filled circles) and $d$ electrons (orange filled circles) for the decoupled
system representing SegP (b), MixP (c) and CheP (d). (e,f) Examples
of the normalized $f$-particle occupations ($N_{f}/L_{y}$) per layer
for tilted potential, lattice $L=20\times20$ and parameters $\gamma=0$,
$U=5$t, and $V=5t$, and $V=10t$ respactively. \label{fig:GS}}
\end{figure}

The phase diagrams of the tilted potential are in general more complex
than for the flat one. Because the tilt is introduced before the coupling
is switched on, the $f$-particle distribution is thermalized already
under the influence of the electrostatic potential $\varepsilon^{\mathrm{tilt}}$.
The interplay of tree different energy scales, namely, the hopping
term $-t$, potential $\varepsilon_{i}^{\mathrm{tilt}}$, and $U$,
leads to a rich ground state summarized in the simplified phase diagram
in Fig.~\ref{fig:GS}(a). The ground-state $f$-particle configurations
are a mixture of segregated \cite{Freericks_seg_1999,LemanskiPRL2002}
and checkerboard phase (CheP)[see Fig.~\ref{fig:GS}(b-d)]. Note that the
phase boundary between pure CheP and mixed phase
(MixP) is stable. We have observed only small changes (below $5\%$)
with increasing system lattice size. On the other hand, the boundary
between pure segregated phase (SegP) and MixP is relevant only for
$L=20\times20$. This boundary shifts to higher $V$ with increasing
lattice size and for $L_{x}\rightarrow\infty$ and finite voltages
the pure SegP might not exist at all.

It was shown before that the spinless FKM at neutral filling ($N_{f}=N_{d}$)
and in the vicinity of the half-filling condition contains different
ordered phases for the same parameters but at different temperatures
\cite{Tran2006,Zonda2012}. The potential tilt leads to a similar,
although less pronounced situation. In Fig.~\ref{fig:Phd}(a) we
show the estimated critical temperatures for the OP-DP transition
(red filled circles) calculated for $V=5t$ and $L=20\times20$. The
segregation starts to form well above this transition. This can be
seen in Fig.~\ref{fig:GS}(e,f), where examples of the normalized
$f$-particle occupation per layer parallel to the interfaces are
plotted for a range of temperatures. The estimated critical temperatures
in Fig.~\ref{fig:Phd}(a) are therefore related to the formation
of the checkerboard pattern in the central region of the lattice.
Consequently, the biggest difference between the phase boundaries
for $\varepsilon^{\mathrm{flat}}$ and $\varepsilon^{\mathrm{tilt}}$
plotted in Fig.~\ref{fig:Phd}(a) is in the parameter regime where
$V\gtrsim U$. In this regime, the checkerboard region is much smaller
than the segregated one. Moreover, because for fixed parameters the
total size of the checkerboard region increases with increasing $L_{x}$
the phase boundary in this region is not saturated yet for $L=20\times20$.
We have observed that with increasing lattice size the boundary approaches
the one for $\varepsilon^{\mathrm{flat}}$. Nevertheless, for a finite
system and tilted potential, the finite voltage can influence the
OP-DP transition even before the coupling is turned on.

\begin{figure}
\includegraphics[width=1\columnwidth]{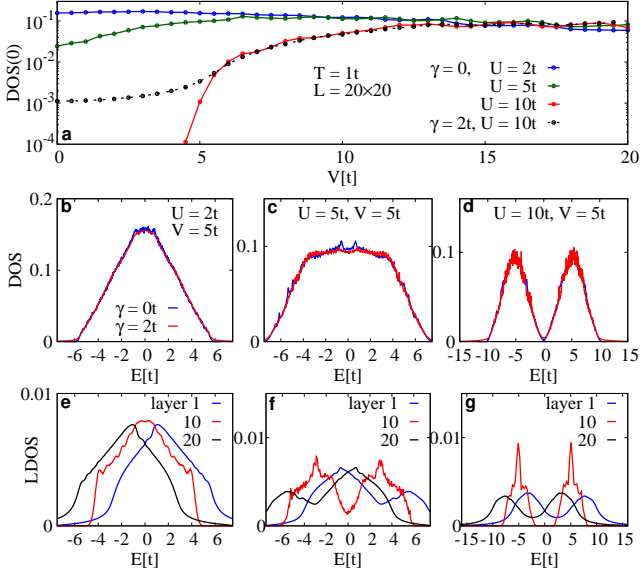}\caption{(a) DOS of $d$-electrons at $E=0$ for a system with tilted potential
calculated as a function of voltage drop at $T=1t$ for a decoupled
($U=2t,5t,10t$) and coupled scenario ($U=10t$, $\gamma=2$). (b-g)
Examples of DOS for tilted potential and different model parameters
calculated for decoupled (blue) and coupled (red) system with $L=20\times20$
. (e-g) Local densities of states calculated for layers at the edges
and in the center of the coupled system. The parameters are
the same as in (b-d). \label{fig:DOStilt}}
\end{figure}

Because of the segregation, the tilt has a dramatic effect on the
system even at high temperatures. This can be seen in the dependence
of the DOS on $U$ for $E=0$ plotted in Fig.~\ref{fig:DOStilt}(a).
A sufficiently tilted potential can close the well-developed Mott
gap even for a decoupled system and strong interaction ($U=10t$ in
Fig.~\ref{fig:DOStilt}(a)). The finite coupling closes the gap in
the entire range of $V$ but the increase of the DOS at the Fermi
level as a consequence of increasing tilt is still visible. We have
observed that introducing the tilt has often a similar effect on the
DOS as has the reduction of $U$ for the flat potential. It closes
the pseudo gap, and the typical sub-CDW gap structures, usually present
for strong interactions even above critical temperature, are merging
with the Hubbard-like satellites (compare Fig.~\ref{fig:DOStilt}(b-d)
with Fig.~\ref{fig:Phd}(c,d)). 

The role of the segregation in this effect even at high temperatures
can be understood by following the examples of the LDOS for particular
system layers in Fig.~\ref{fig:DOStilt}(e-g). Fig.~\ref{fig:DOStilt}(e)
shows the LDOS of the coupled system for $U=2t$ and $V=5t$, where
the ground state is completely segregated. All LDOSs resemble the DOS
of a weakly interacting system, where the absence or presence of $f$ particles
at the interfaces manifests itself by a constant shift competing with
$\varepsilon_{i}^{\mathrm{tilt}}$. For $U=2t$ and $V=5t$ (Fig.~\ref{fig:DOStilt}(f)),
the LDOS in the center region has a profile typical for the disordered
phase at intermediate coupling, but the effect of the segregation
is still obvious at the interfaces. As there is no segregation for
$U=10t$ and $V=5t$ the ``center'' of $\rho_{L}(E)$ and
$\rho_{L}(E)$, respectively, and the system LDOS at interfaces are almost aligned
($\epsilon_{L}\cong\varepsilon_{1}^{\mathrm{tilt}}+U/2$ and $\varepsilon_{L_{x}}^{\mathrm{tilt}}+U/2\cong\epsilon_{R}$),
therefore LDOSs are shifted mostly by $\varepsilon^{\mathrm{tilt}}$
and otherwise resemble a broadened DOS of the strongly interacting
disordered phase.

\subsubsection{Transport properties\label{subsec:TransportT}}

\begin{figure}
\includegraphics[width=1\columnwidth]{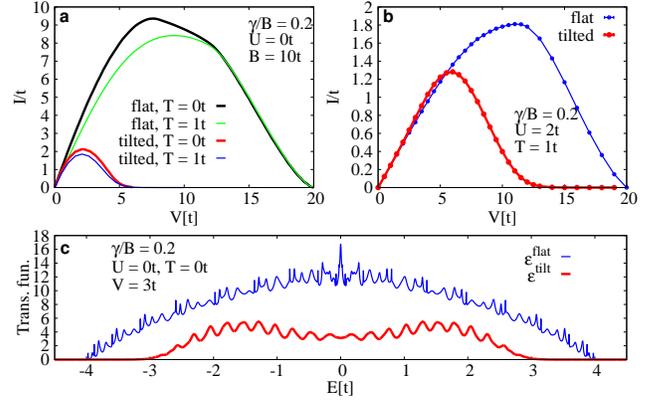}\caption{(a-b) Comparisons of current - voltage characteristics calculated for
flat and tilted potential at high temperature $T=1t$ and $0t$.
The left figure shows the noninteracting case, and the right one was
calculated for $U=2t$. The other parameters are $\gamma=2t$, $B=10t$,
and $L=20\times20$. (c) Comparison of the transmission functions for
the noninteracting system with flat and tilted potential for small
voltage $V=3t$. \label{fig:FlatVsDrop}}
\end{figure}

As already shown above, the tilt of the potential can have a profound
effect on the electron distribution. A similar result was obtained
for the Hubbard model by Okamoto \cite{Okamoto2008}. However, in
that particular study both potential profiles still gave rather
similar current-voltage characteristics. In the case considered here,
the consequences of the tilted potential for the charge transport
are significant. This can be seen already for the noninteracting
case plotted in Fig.~\ref{fig:FlatVsDrop}(a), where we compare $I-V$
characteristics of both potential profiles at high and zero temperature.
The current for the tilted case is significantly smaller than for
the flat potential. This is also signaled by a significant decrease
of the transmission function (Fig.~\ref{fig:FlatVsDrop}(c)) already
for small voltages. Although a clear difference between tilted and
flat geometry is evident for the interacting system as well (see Fig.~\ref{fig:FlatVsDrop}(b)),
it is much less dramatic than for $U=0$. This may partially explain
the similarities in the currents in Okamoto's work\cite{Okamoto2008},
where $U$ was rather large.

\begin{figure}
\includegraphics[width=0.75\columnwidth]{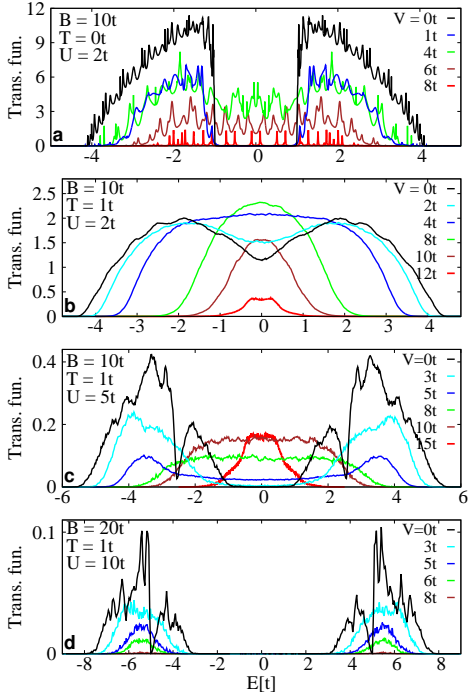}\caption{(a-d) Transmission functions for the tilted system potential and different
values of $U$, $T$, and $V$. Other parameters are: lattice size
$L=20\times20$, coupling $\gamma=0.2B$, and band half width: (a-c)
$B=10t$ and d) $B=20t$. \label{fig:TransDrop}}
\end{figure}

The significant difference of the transport results for $\varepsilon^{\mathrm{flat}}$
and $\varepsilon^{\mathrm{tilt}}$ can be understood by comparing
the transmission functions. Whereas the transmission functions for
the flat potential in Fig.~\ref{fig:TransFlat} do not change significantly
with voltage unless $V>B$, the tilt of the potential has a profound
effect already for small voltages (see Fig.~\ref{fig:TransDrop}).
This is further complicated by the strong spatial dependence of the
LDOS on $V$ and $U$ that influences the overall transmission differently
in central and segregated regions. The low transmission at energies
$\left|E\right|<U/2$ for the central region is responsible for the
gap in the transmission function but it has to compete with growing
segregation that is closing the gap. 

The transmission function changes rapidly with increasing voltage
for $U=2t$ at both zero and high temperatures (Fig.~\ref{fig:TransDrop}(a,b)).
First, the pseudo-CDW gap closes. This again resembles the effect
of lowering $U$. Besides filling the gap, the increasing voltage
leads to narrowing of the transmission function. The combination of
these two effects at high temperature leads initially to a significant
increase of the transmission function around $E=0$. Further increase
of the voltage introduces the segregation in the $f$-particle subsystem
and the transmission function drops down. Consequently, when we compare
the $I-V$ characteristics of the flat and tilted potential, plotted
in Fig.~\ref{fig:FlatVsDrop}(b), the curves coincide only for $V<U$.
The tilt shifts the maximum of the current to lower voltage ($V\sim5.5t$)
and a rather steep decline follows for the voltages that are well
below $V\sim B$. Therefore, and in contrast to the flat potential,
the drop of the current is a consequence of the vanishing transmission
function and not of closing the band window $W$.

\begin{figure}
\includegraphics[width=1\columnwidth]{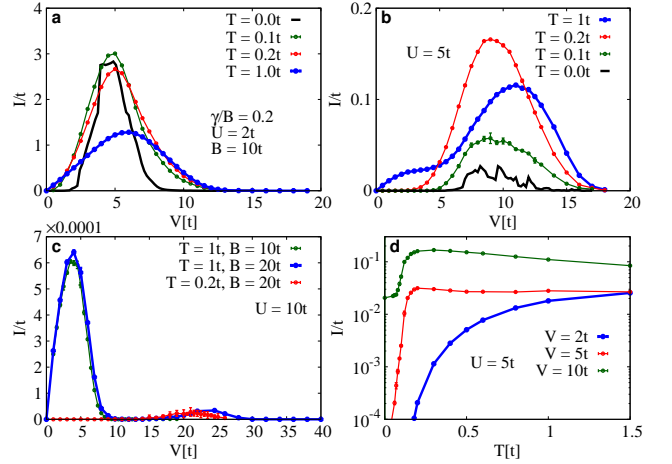}

\caption{The current-voltage relation for tilted system potential, system size
$L=20\times20$, coupling $\gamma/B=0.2$ and parameters (a) $U=2t$,
$B=10t$; (b) $U=5t$, $B=10t$; and (c) $U=10t$, $B=10t$ and $20t$
and different temperatures. \label{fig:CurrentDrop} (d) The transmission
function at $E=0$ as a function of temperature for $V=5t$ and various
values of $U$.}
\end{figure}

The situation is more complicated for $U=5t$. On one hand, the increasing
voltage is closing the gap, but, on the other, it pushes the in-gap
maxima and Hubbard-like satellites together. This makes the pseudo
gap shallower but also effectively broader. As for the weak interaction,
with increasing $V$ the transmission function initially increases
in the central region (Fig.~\ref{fig:TransDrop}(c)). The transmission
function drops again at higher voltages where the segregation starts
to form. The voltage dependence of the transmission function for $U=10t$
is similar to $U=5t$ (Fig.~\ref{fig:TransDrop}(d)) but the central
region is suppressed which is related to the gap in the LDOS of the
central layers even for high voltages as seen in Fig.~\ref{fig:DOStilt}(g). 

The complicated dependences of the transmission function on the voltage
for $U=5t$ and $U=10t$ is imprinted in the $I-V$ characteristics
plotted in Fig.~\ref{fig:CurrentDrop}(b,c). The current for $U=5t$
and $T=1t$ has almost a plateau for $1\lesssim V<5$. This is because
two opposite tendencies are active in this region. The increasing
width of the Fermi window should increase the current, but it has
to compete with the effective increase of the width of the pseudo gap
and the decreasing amplitude of the Hubbard-like satellites. The following
maximum reflects the increase of the transmission function in the
central region (Fig.~\ref{fig:TransDrop}(c)). This effect is even
more pronounced for strong interaction, resulting in two tiny (see
the scale) maxima in the current.

The kinks and sharp steps visible in the $I-V$ characteristics at
zero temperature in Fig.~\ref{fig:CurrentDrop}(a,b) reflect the
changes in the ground-state $f$-particle distribution. The MixP does
not change from CheP to SegP in a continuous way but rather steplike.
We have found that with increasing $V$ the segregated phase always
grows by filling and emptying a whole layer. Every such change can
lead to a sharp feature in the $I-V$ characteristics.

The presence of the segregated regions significantly influences the
temperature dependence of the current as illustrated for $U=5t$ in
Fig.~\ref{fig:CurrentDrop}(d). We have not observed any sharp increase
of the current below critical temperature typical for high-voltage
dependencies for the flat potential. This is a consequence of the
significant suppression of the Hubbard-like satellites for increasing
voltage. A sharp exponential decrease of the current below the critical
temperature can be observed for $V=5t$. With further lowering of
the voltage, the sharp steplike decrease changes into a smooth dependence
typical for the DPs-OP transition.

\subsubsection{System size effects\label{subsec:SSET}}

\begin{figure}
\includegraphics[width=1\columnwidth]{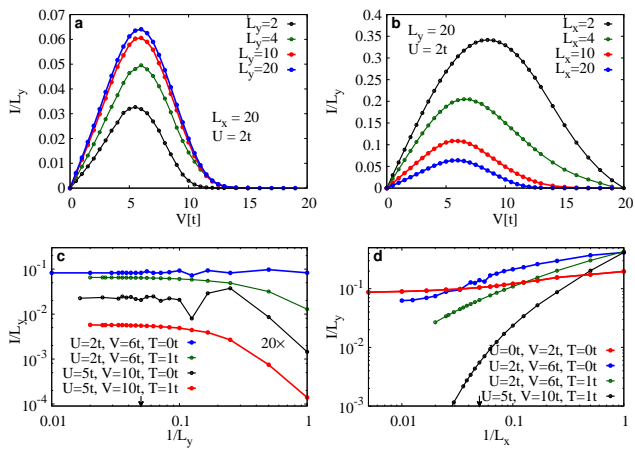}\caption{Current densities plotted: (a) as a function of voltage for a system
with $L_{x}=20$ and different vertical sizes $L_{y}$ at $T=1t$; (b)
as a function of voltage for a system with $L_{y}=20$ and different numbers
of layers $L_{x}$ at $T=1t$; (c) as a function of inverse linear size
$1/L_{y}$ for system with $L_{x}=20$ and optimal $V$, where the data for
$U=5t$ and $T=1t$ were multiplied by $20$ for the sake of clarity; and (d)
as a function of inverse linear size $1/L_{x}$ for a system with $L_{y}=20$.
\label{fig:FinSizEfDrop}}
\end{figure}

The finite-size dependence for the system with tilted potential is
quite complicated. This is illustrated in Fig.~\ref{fig:FinSizEfDrop}.
As for the flat potential case, the current density converges rapidly
with increasing $L_{y}$ and is practically saturated for $L_{y}=20$.
The saturation is monotonous for high temperature, but it oscillates
at $T=0t$. Small oscillations are evident even above $L_{y}=20$.
The potential tilt has a bigger influence on the scaling in the $x$ direction.
The red filled circles in Fig.~\ref{fig:FinSizEfDrop}(d) show that
this is true already for the noninteracting case, where the $f$ particles
do not play a role. The current density seems to be converging to
finite values for weakly and noninteracting cases at zero temperature.
However, this convergence is slow and much larger lattices than for
the flat potential are needed to confirm the plateau. For the same
reason, it is not obvious from the numerical results that the weakly
interacting case at high-temperature approaches zero for $L_{x}\rightarrow\infty$.
All this shows, that the profile of the system potential has a dramatic
effect on the steady-state transport. 

\section{Conclusions \label{sec:Conclusions}}

We have studied charge transport in a model heterostructure, where
the central system is described by a Falicov-Kimball model, which
is coupled to metallic leads. The method used is based on a combination
of a sign-problem-free MC approach and nonequilibrium Green's-functions
techniques. Due to its efficiency, this methodology allows us to study
large enough structures to approach the condensed phase limit.

We have specifically considered charge transport through a 2D FKM
for two types of electrostatic potentials in the system, including
a flat and a tilted form. In both cases, the transport characteristic
is closely related to the phases of the FKM. For a flat potential,
the current signals a DP-OP transition by a sharp steplike increase
or decrease of the current depending on the bias voltage. The transition
from DPw to DPs can be clearly identified from the properties of the
transmission function and it is not affected even by a high voltage.
Different regimes of localization can be seen even for high voltages.
The tilted potential, on the other hand, introduces new phases including
charge segregation. A significant tilt can close the Mott gap and
fill in the CDW pseudo gap and therefore significantly affects the
current-voltage characteristics.

We have, furthermore, analyzed the dependence of the transport properties
on the size of the system and shown that a size of $L_{x}=20\times20$
is sufficient for addressing the condensed phase limit. Moreover,
we have found that the finite-size scaling of the current is qualitatively
different for weak and strong interaction.

In the present paper, we have adopted several simplifications, the
lifting of which could be interesting for future investigations. First,
the leads were modeled by mutually disconnected semi-infinite chains.
Second, it would be interesting to include part of the leads at the
interfaces into the system, because the proximity to the system can
change the electron distribution in these areas of the leads \cite{Freericks2001,Okamoto_Milis_2004}.
Finally, we have assumed that in equilibrium the system and the leads
have the same effective chemical potential and, therefore, there is
no evident effect of electronic charge reconstruction at the interfaces
\cite{Okamoto_Milis_Nature2004}. This will typically not be the case
in real materials. The extension of the paper to include these aspects
will the subject of future work.
\begin{acknowledgments}
The authors acknowledge support by the state of Baden-Württemberg
through bwHPC (High Performance Computing in Baden-Württemberg) and the German Research Foundation (Deutsche Forschungsgemeinschaft) through grant
no INST 40/467-1 FUGG. We thank Rainer H\"artle for inspiring and
very helpful discussions. 
\end{acknowledgments}


%

\end{document}